\documentclass[iop, twocolumn]{openjournal}

\usepackage{graphicx}	
\usepackage{float}
\usepackage{amsmath}	
\usepackage{dcolumn}
\usepackage{bm}
\usepackage{orcidlink}
\usepackage[dvipsnames]{xcolor} 
\usepackage{textgreek}
\usepackage[utf8]{inputenc}
\usepackage[english]{babel}
\usepackage{lastpage}
\usepackage{hyperref}
\hypersetup{
    unicode, 
    colorlinks=true,
    linkcolor=linkcolor,
    citecolor=linkcolor,
    filecolor=linkcolor,
    urlcolor=linkcolor,
}
\usepackage{color,colortbl}
\definecolor{linkcolor}{rgb}{0.0,0.3,0.5}
\usepackage{tensind}
\tensordelimiter{?}
\DeclareGraphicsExtensions{.bmp,.png,.jpg,.pdf}

\usepackage{verbatim}
\usepackage[normalem]{ulem}
\usepackage{soul}

\newcommand{\bna}{\bm \nabla}
\newcommand{\bal}{\bm \alpha}
\newcommand{\be}{\begin{equation}}
\newcommand{\ee}{\end{equation}}
\newcommand{\ba}{\begin{eqnarray}}
\newcommand{\ea}{\end{eqnarray}}

\newcommand{\de}{\delta}

\newcommand{\uu}{\mathbf{u}}

\begin{document}

\title{Neural posterior estimation of the line-of-sight and subhalo populations in galaxy-scale strong lensing systems}
\shorttitle{Neural Inference of Lensing Substructure}
\shortauthors{Dhanasingham, Cyr-Racine, and Gilman}

\author{\vspace{-1.0cm}Birendra Dhanasingham\,\orcidlink{0000-0003-0203-3853}}
\email{bdhana@umn.edu}
\affiliation{Minnesota Institute for Astrophysics, University of Minnesota, 116 Church St SE, Minneapolis, MN 55455, USA}
\affiliation{School of Physics and Astronomy, University of Minnesota, 116 Church St SE, Minneapolis, MN 55455, USA}

\author{Francis-Yan Cyr-Racine\,\orcidlink{0000-0002-7939-2988}}
\email{fycr@unm.edu}
\affiliation{Department of Physics and Astronomy, University of New Mexico, 210 Yale Blvd NE, Albuquerque, NM 87106, USA}

\author{Daniel Gilman\orcidlink{0000-0002-5116-7287}}
\email{gilmanda@uchicago.edu}
\affiliation{Department of Astronomy and Astrophysics, University of Chicago, 5640 S Ellis Ave, Chicago, IL 60637, USA}
\affiliation{Brinson Prize Fellow}

\begin{abstract}

Strong gravitational lensing is a powerful probe for studying the fundamental properties of dark matter on sub-galactic scales. Detailed analyses of galaxy-scale lenses have revealed localized gravitational perturbations beyond the smooth mass distribution of the main lens galaxy, largely attributed to dark matter subhalos and intervening line-of-sight halos. Recent studies suggest that, in contrast to subhalos, line-of-sight halos imprint distinct anisotropic features on the two-point correlation function of the effective lensing deflection field. These anisotropies are particularly sensitive to the collisional nature of dark matter, offering a potential means to test alternatives to the cold dark matter paradigm. In this study, we explore whether a neural density estimator can directly identify such anisotropic signatures from galaxy-galaxy strong lens images. We model the multipoles of the two-point function using a power-law parameterization and train a neural density estimator to predict the corresponding posterior distribution of lensing parameters, alongside parameter distributions for dark matter substructure. Our results show that recovering the dark matter substructure mass functions and mass-concentration parameters remains challenging, owing to difficulties in generating uniform training data set while using physically motivated priors. We also unveil an important degeneracy between the line-of-sight halo mass-function amplitude and the subhalo mass-function normalization. Furthermore, the network exhibits limited accuracy in predicting the two-point function multipole parameters, suggesting that both the training data and the adopted power-law fitting function may inadequately represent the true underlying structure of the anisotropic signal.


\end{abstract}

\maketitle



\section{Introduction}

Dark matter plays a pivotal role in shaping the intricate structure we observe across the universe today. According to the standard Lambda Cold Dark Matter ($\Lambda$CDM) model, the formation of dark matter structures follows a hierarchical, bottom-up process, with smaller halos forming first and merging to form larger structures with increasing mass and size. While smaller dark matter halos may lack the gravitational pull to retain gases and remain unseen, larger halos are known to host visible structures such as clusters, groups, and galaxies. Although the $\Lambda$CDM model has been highly successful in explaining the formation and evolution of structures on mass scales larger than $\sim10^{11}\,{M}_\odot$ and length scales larger than $\sim 1$ Mpc, questions persist regarding its efficacy in the non-linear regime, where small-scale halos play a significant role \citep{Bullock:2017xww}. Various alternative dark matter models have emerged over time to address the shortcomings of the $\Lambda$CDM model in this regime. For instance, warm dark matter (WDM) models \citep{Bond_1983, Bode:2000gq, Dalcanton:2000hn, Schneider:2013ria, Viel_2013, Benson_2013, Pullen_2014, Lovell_2020} propose modifications to linear theory predictions within the $\Lambda$CDM framework, while self-interacting dark matter (SIDM) scenarios \citep{Spergel_2000, Tulin:2012wi, Tulin:2013teo, Rocha_2013, Peter_2013, Kaplinghat_2013, Kaplinghat_2014_1, Kaplinghat_2014_2, Kaplinghat:2015aga, Elbert_2015, Burger_2019} alter nonlinear predictions within collisionless CDM models described by the Boltzmann equation. Consequently, the precise characterization of dark matter halos at sub-galactic scales offers a critical assessment of the $\Lambda$CDM paradigm.

Strong gravitational lensing serves as a potent tool for constraining dark matter microphysics, particularly in this non-linear regime. In addition to the gravitational effects induced by the main lensing galaxy's smooth mass distribution, the presence of dark matter substructures introduces subtle distortions in the observed images \citep{Bolton_2006, Bolton_2008, Gavazzi_2008, Auger_2009, Brownstein_2012, Shu_2016, oldham_2017, Cornachione_2018, DES:2019mte}. A meticulous examination of these distortions offers valuable insights into the underlying physics governing these substructures. Recent observations of galaxy-galaxy strong lensing have resulted in several reported detections of dark matter halos \citep{vegetti2010a, vegetti2010b, vegetti2012, vegetti2014, Nierenberg_2014, Hezaveh2016, Nierenberg_2017,Despali:2024ihn,Stacey:2025vth,Tajalli:2025qjx,Powell:2025rmj,McKean:2025ozb}. These halos may either represent subhalos within the main lens \citep{Mao:1998aa, Chiba:aa} or halos along the line of sight that coincidentally project near the lensed images \citep{Keeton:2003aa, Xu:2011ru, Li:2016afu, Despali:2017ksx, Amorisco_2021}. Current research efforts aimed at constraining dark matter through strong lensing predominantly involve either modeling individual substructures within lens systems \citep[see e.g.,][]{vegetti2010a, vegetti2010b, vegetti2012, vegetti2014, Hezaveh2016, Ritondale:2018cvp, Minor_2021, Sengul:2021lxe,Sengul:2022edu,Zhang:2022djp} or considering the collective impact of substructure on lensing images to derive statistical constraints \citep[see e.g.,][]{Xu:2011ru, Daylan:2017kfh, DiazRivero:2017xkd, DiazRivero:2018oxk,Bayer:2018vhy,Cyr-Racine_2019, Gilman:2019vca, Gilman:2019nap, Gilman:2019bdm, Hogg_2023, Keeley:2023sad, Dhanasingham:2022btr, Dhanasingham:2023thg,Dhanasingham:2022nox,Bayer_2023,Gilman:2025fhy,Keeley:2025oig}.

In recent years, machine learning has emerged as a valuable tool for analyzing strong lens images and effectively constraining dark matter substructure within strong lens systems using observational data. Numerous studies have demonstrated that trained neural networks can robustly constrain the parameters of strong lens models directly from observations \citep[see e.g.,][]{Hezaveh_2017, Perreault_Levasseur_2017, 2021A&A...646A.126S, Wagner-Carena_2021,Wagner-Carena:2024axc,Filipp:2024yef,Barco:2024vpe,Barco:2025imh}. Additionally, within the strong lens community, there has been significant interest in automated modeling of strong gravitational lenses with neural networks \citep[see e.g.,][]{2023MNRAS.522.5442G}, as well as the development of machine learning-based tools to identify gravitational lenses from upcoming surveys \citep[
see e.g.,][]{Schaefer_2018, Cheng_2020, Li_2020, Wilde_2022, Rezaei_2022, Rojas_2023, 2025MNRAS.538.1081R}. Furthermore, while many studies have developed tools to successfully constrain the subhalo mass function using strong lens observations \citep[see e.g.,][]{Brewer:2015yya, Brehmer:2019jyt, Ostdiek:2020mvo, Montel:2022fhv, Wagner-Carena:2022mrn}, others have focused on detecting dark matter substructure in mock strong lens images\citep[see e.g.,][]{Lin:2020nrc, DiazRivero:2019hxf, Khosa:2019qgp, Tsang:2024agc}. 

With the inventory of strong lenses available for study expected to expand significantly, with over a thousand identified lenses currently and tens of thousands anticipated through upcoming wide-field optical imaging surveys \citep{2009arXiv0912.0201L, 2011arXiv1110.3193L, Serjeant_2014, Collett_2015, Weiner_2020, Mao_2022}, machine learning-based approaches offer the advantage of statistical inference to better handle these large datasets and impose tight constraints on dark matter properties. In \cite{Wagner-Carena:2022mrn}, a simulation-based inference (SBI) pipeline was employed, utilizing a trained neural network to estimate the posterior density of the subhalo mass function and to place constraints on lens populations. However, they observed only a weak correlation between the actual and inferred normalization of the subhalo mass function. They concluded that by integrating their network with a hierarchical inference framework, it is possible to accurately infer the subhalo mass function across various configurations while efficiently scaling to populations with hundreds of lenses. Although these findings are encouraging, neural posterior estimators still have limitations in terms of their constraining power. In \cite{Wagner-Carena:2024axc}, the primary methodological limitation of neural posterior estimation was identified as the size of the training dataset. \cite{Wagner-Carena:2024axc} then utilized sequential neural posterior estimation to significantly enhance the quality of the training samples and thus improve the constraining power of the data. 

In their work, \cite{Dhanasingham:2022nox} introduced the effective multiplane gravitational lensing framework, which considers the collective impact of multiple lens planes in a strong lens system. They highlighted that while subhalos within the main lens produce fluctuations that are largely statistically isotropic and homogeneous, dark matter halos along the line-of-sight exhibit distinct anisotropic signatures in convergence maps, which suggest a distinct set of perturbative features relative to single-plane lensing. These signatures manifest as a detectable, non-zero parity-even quadrupole moment in the two-point correlation function of the effective convergence field, and as non-zero curl components in the effective deflection field that exhibit a parity-odd quadrupole moment in the two-point function, once these two-point correlation functions are decomposed onto a set of orthonormal basis functions. While the line-of-sight dark matter halos contribute exclusively to the parity-odd and parity-even quadrupole moments, the monopole moment of the two-point correlation function of the effective convergence field arises from the combined contribution of subhalos in the main lens plane and halos along the line-of-sight. These multipole moments offer a statistical means to differentiate between line-of-sight halos and main lens dark matter subhalos. Building on this, \cite{Dhanasingham:2023thg} demonstrated the sensitivity of these multipole moments to both the amplitude and velocity dependence of dark matter self-interaction cross-sections. In this study, we evaluate the potential of extracting these multipole moments from strong lensing observations using a machine learning approach. Leveraging the simulation-based inference pipeline \textsc {paltas}\footnote{https://github.com/swagnercarena/paltas} developed by \cite{Wagner-Carena:2022mrn}, we train a neural posterior estimator of the two-point function multipoles by statistically analyzing the impact of substructure on strong lens images.

This paper is organized as follows: Section~\ref{data_gen} details the generation of training and validation datasets, along with the computation of two-point function statistics. Section~\ref{SBI} presents a concise overview of the Simulation-Based Inference (SBI) framework utilizing Convolutional Neural Networks (CNNs). In Section~\ref{Results_4}, we assess the effectiveness of a trained neural posterior density estimator in extracting two-point correlation function multipole statistics and inferring parameters that characterize the properties of dark matter substructure. Finally, Section~\ref{Discussion} summarizes results of our study and explores their broader implications.

\section{Data generation} \label{data_gen}

In this section, we delineate the process of generating training and validation datasets, which encompasses the simulating of strong lensing images and the computation of two-point correlation function statistics. These datasets are utilized to train our neural networks and subsequently evaluate our model predictions against mock images post-training.

\subsection{Simulating strong lens systems}\label{lens_image_sim}

In this study, we aim to generate simulated strong lens images that closely resemble the quality of observed images captured by the \textit{Hubble Space Telescope} (\textit{HST}). To achieve this, we utilize the open-source lens modeling \textsc{python} software package \textsc {lenstronomy}\footnote{https://github.com/lenstronomy} \citep{Birrer:2018xgm, Birrer2021}. Additionally, we employ \textsc {pyhalo}\footnote{https://github.com/dangilman/pyHalo} \citep{Gilman:2019nap} to incorporate dark matter subhalos and line-of-sight halos into the strong lens systems, which contribute to perturbations. Throughout our work, we adopt a flat $\rm \Lambda CDM$ cosmology, following the Planck 2018 results \citep{planck_2018}. Producing a lensed image involves several components: a main lens galaxy (macrolens), a population of dark matter subhalos associated with the main lens galaxy, line-of-sight dark matter halos, a source galaxy, and specifications for the instrumental setup. Below, we elaborate on our selections for each of these elements to create training and validation simulated images.

\begingroup
\setlength{\tabcolsep}{10pt}
\renewcommand{\arraystretch}{1.8}
\begin{table*} 
	\centering
	\caption{Parameters of the primary elements in a strong gravitational lensing system, along with the distributions specific to each, employed for generating training, validation, and test mock images in our study.}
	\begin{tabular}{llr} 
		\hline\hline
		\textbf{Parameter} & \textbf{Definition}  &\textbf{Distribution} \\
		\hline \hline
  \textbf{EPL Main Lens}\\
  $\theta_{\rm E}[{\rm arcsec}]$ & Einstein Radius & $\mathcal{U}(0.8, 1.2)$\\
  $\gamma_{\rm macro}$ & Logarithmic slope of main lens power-law mass model & $\mathcal{U}(1.9, 2.1)$\\
  $q_{\rm e}$ & Ratio of minor axis to major axis & $\mathcal{U}(0.25, 0.90)$\\
  $\phi_{\rm e}[{\rm radians}]$ & Orientation angle of ellipticity & $\mathcal{U}(-\pi, \pi)$ \\
  $(x_{\rm macro}, y_{\rm macro})[{\rm arcsec}]$ & $x-$ and $y-$ coordinates of the main lens & $(0,0)$\\
  $\log_{10}(M_{\rm halo}[{M}_\odot])$ & Host halo mass & 13.3 \\
  $z_{\rm lens}$ & Main lens redshift & 0.5\\
  \hline
  \textbf{External Shear}\\
    $\log_{10}\gamma$ & External shear strength & $\mathcal{U}(-2.5, -1.0)$ \\
  $\phi_{\rm \gamma}[{\rm radians}]$ & External shear angle & $\mathcal{U}(-\pi, \pi)$\\
  		\hline
  \textbf{Main Lens Light Profile} \\
  $I_{\rm 0}$ & Amplitude value at the half-light radius & $\mathcal{U}(1.0, 20.0)$\\
  $R_{\rm S\acute{e}rsic} [{\rm arcsec}]$ & Half-light radius & $\mathcal{U}(0.05, 0.5)$ \\
  $n_{\rm S\acute{e}rsic}$ & S\'{e}rsic index & $\mathcal{U}(2.0, 6.0)$ \\
  $q_{\rm e, S\acute{e}rsic}$ & Ratio of minor axis to major axis & $\mathcal{U}(0.5, 0.99999)$\\
  $\phi_{\rm e, S\acute{e}rsic}[{\rm radians}]$ & Orientation angle of ellipticity & $\mathcal{U}(-\pi, \pi)$ \\
  $(x_{\rm S\acute{e}rsic}, y_{\rm S\acute{e}rsic})[{\rm arcsec}]$ & $x-$ and $y-$ coordinates of the main lens light profile & $(0,0)$ \\
  \hline
  \textbf{Dark Matter Substructure}\\
  $\log_{10}(m_{\rm min}[{M}_\odot])$ & Minimum dark matter halo mass to render & 6.0\\
  $\log_{10}(m_{\rm max}[{M}_\odot])$ & Maximum dark matter halo mass to render & 10.0\\
  $\log_{10}(\Sigma_{\rm sub}[\rm{kpc}^{-2}])$ & Subhalo mass function normalization & $\mathcal{U}(-3.0,-1.0)$\\
  $\alpha$ & Logarithmic slope of the subhalo mass function & -1.9\\
  $\log_{10}(m_0[{M}_\odot])$ & Subhalo power-law pivot mass & 8.0\\
  $\de_{\rm LOS}$ & Line-of-sight halo mass function normalization & $\mathcal{U}(0.0, 2.0)$\\
  $\theta_{\rm cone}[{\rm arcsec}]$ & Opening angle of the double cone rendering volume & 8.0\\
  $\log_{10} \beta$ & Free overall scaling factor of the mass-concentration relationship & $\mathcal{U}(-1.0, 1.0)$ \\
		\hline 
	\end{tabular}
\end{table*}\label{params}
\endgroup

\subsubsection{Main lens}

We represent the main lens using an elliptical power-law (EPL) profile \citep{1994A&A...284..285K, 1998ApJ...502..531B, Tessore_2015}, where the convergence is parameterized as 
\be 
\kappa_{\rm EPL}(x,y)=\frac{3-\gamma_{\rm macro}}{2}\left(\frac{\theta_{\rm E}}{\sqrt{q_{\rm e}x^2+\frac{y^2}{q_{\rm e}}}} \right)^{\gamma_{\rm macro}-1}.
\ee Here, $1 < \gamma_{\rm macro} < 3$ denotes the power-law slope of the mass distribution, $\theta_{\rm E}$ is the Einstein radius, and $0 < q_{\rm e} \leq 1$ is the ratio of the minor and major axes. Given a specific orientation angle of ellipticity, $\phi_{\rm e}$ and axis ratio $q_{\rm e}$, the eccentricity components $e_1$ and $e_2$ can be expressed as 
\be \label{Eq_e1}
e_1 = \frac{1 - q_{\rm e}}{1 + q_{\rm e}} \cos(2\phi_{\rm e}) 
\ee
and
\be \label{Eq_e2}
e_2 = \frac{1 - q_{\rm e}}{1 + q_{\rm e}} \sin(2\phi_{\rm e}).
\ee
The external gravitational shear field, induced by matter surrounding the main lens, is defined by its strength $\gamma$ and orientation angle $\phi_\gamma$, represented by the coordinate values $\gamma_1 = \gamma \cos(2\phi_\gamma)$ and $\gamma_2 = \gamma \sin(2\phi_\gamma)$ in their respective space.

In our simulated strong lensed images, the light distribution of the lens galaxy follows an elliptical S\'{e}rsic profile \citep{1963BAAA....6...41S}, which can be described as

\be
 I(R) = I_0 \exp \left\{ -b_{n_{\rm S\acute{e}rsic}} \left[\left(\frac{R}{R_{\rm S\acute{e}rsic}}\right)^{\frac{1}{n_{\rm S\acute{e}rsic}}}-1\right]\right\},
\ee where $R=\sqrt{q_{\rm e, S\acute{e}rsic}x^2+\frac{y^2}{q_{\rm e, S\acute{e}rsic}}}$ is the distance from the center, $R_{\rm S\acute{e}rsic}$ represents the half-light radius, $I_0$ is the amplitude at $R_{\rm S\acute{e}rsic}$, and $n_{\rm S\acute{e}rsic}$ is the S\'{e}rsic index. Drawing upon the analytical formulations proposed by \cite{1989woga.conf..208C} for estimating $b_{n_{\rm S\acute{e}rsic}}$, we employ the expression $b_{n_{\rm S\acute{e}rsic}}\approx 1.999n_{\rm S\acute{e}rsic}-0.327$ for $0.5< n_{\rm S\acute{e}rsic} <10$. We use equations~\eqref{Eq_e1} and~\eqref{Eq_e2} to sample the eccentricity components of the light profiles across given distributions of the axis ratio $ q_{\rm e, S\acute{e}rsic} $ and orientation angle $ \phi_{\rm e, S\acute{e}rsic} $. As a simplifying assumption, we posit that the centroid of the main lens light coincides with the center of the main lens mass profile. The permissible ranges of variation for the main lens light profile parameters are delineated in Table~\ref{params}.

\subsubsection{Subhalos and line-of-sight halos}
\begin{figure*}
\begin{center}
	\includegraphics[clip, trim=0cm 0cm 0cm 0cm, width=1\textwidth]{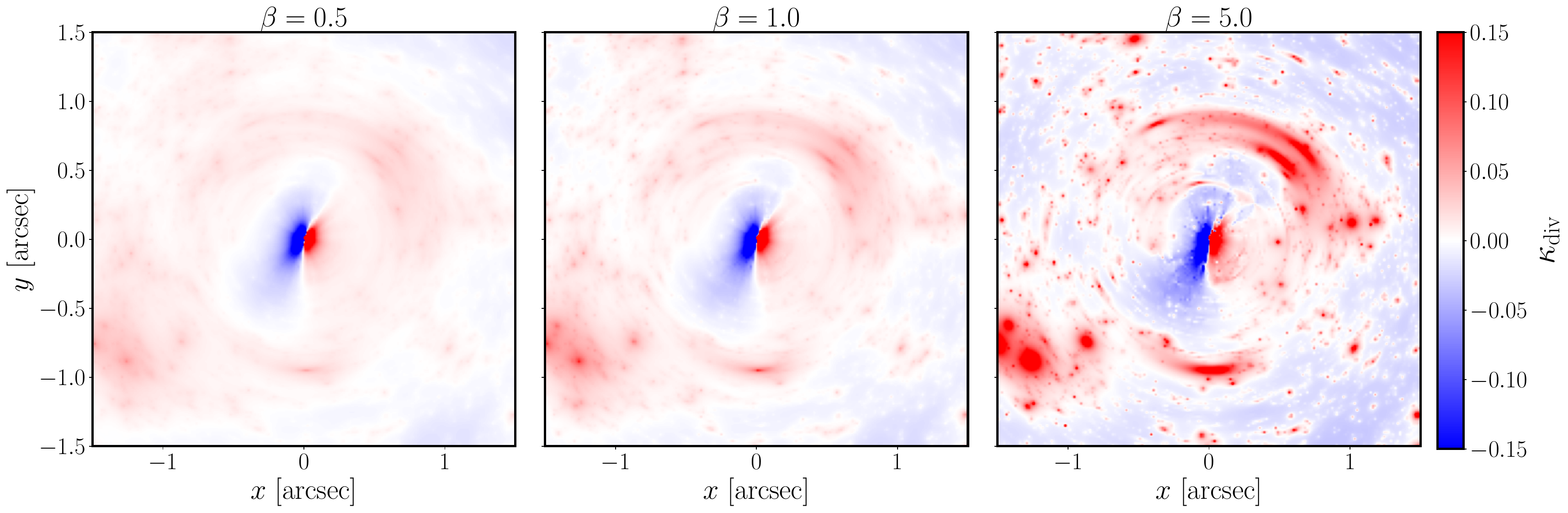}
\end{center}	
\caption{\label{fig:kappa_maps}Effective multi-plane convergence maps ($\kappa_{\rm div}$) for the same CDM halo realization, shown for three different values of the rescaling factor $\beta$ controlling the amplitude of the mass–concentration relationship.}
\end{figure*}

The subhalos within the parent dark matter halo and the dark matter halos along the line-of-sight between the observer and the source are represented using the truncated Navarro--Frenk--White (tNFW) profile \citep{tNFW_2009}, which is parameterized as
\be
\rho(r) = \rho_s\left(\frac{\tau^2}{x(1+x)^2(x^2+\tau^2)} \right)
\ee with $x=r/r_{\rm s}$ and $\tau=r_{\rm t}/r_{\rm s}$ with scale radius $r_{\rm s}$ and truncation radius $r_{\rm t}$. We calculate $r_{\rm s}$ and $\rho_{\rm s}$ for a halo, defined with a mass $M_{200}$, relative to the critical density of the Universe at the halo's redshift, utilizing the mass-concentration relationship outlined by \cite{Diemer:2018vmz}, considering a scatter of 0.2 dex \citep{Wang:2020hpl}. We further adjust the amplitude of the mass-concentration relationship by introducing a free overall scaling factor, $\beta$, which is randomly selected from a distribution described in Table \ref{params}. Fig.~\ref{fig:kappa_maps} illustrates how varying $\beta$ influences the effective multi-plane convergence maps ($\kappa_{\rm div}$) for a single CDM halo realization. The subhalos undergoes tidal truncation based on their three-dimensional positions within the parent dark matter halo, modeled using a Roche-limit approximation that assumes an approximately isothermal global mass profile \citep{Cyr-Racine:2015jwa, Gilman:2019nap}. Similarly, we implement tidal truncation to line-of-sight halos at $r_{50}$, which is comparable to the halo splashback radius \citep{More:2015ufa}.

The subhalos are then distributed according to a power-law mass function represented by
\be \label{sub_m_f}
\frac{{\rm d}^2N_{\rm sub}}{{\rm d}m{\rm d}A}= \frac{\Sigma_{\rm sub}}{m_0}\left(\frac{m}{m_0} \right)^\alpha\mathcal{F}(M_{\rm halo}, z).
\ee with a pivot mass $m_0=10^{8}{M_\odot}$. Here 
\be \label{sub_m_f_F}
\log_{10}(\mathcal{F}) = k_1 \log_{10}\left(\frac{M_{\rm halo}}{10^{13} M_\odot}\right)+k_2 \log_{10}(z+0.5)
\ee denotes the scaling function that dictates the differential projected number density of the subhalos as a function of the host halo mass $M_{\rm halo}$ and the redshift $z$. The host scaling factor ($k_1$) and redshift scaling factor ($k_2$) are assigned values of 0.88 and 1.7, respectively \citep{Gilman:2019bdm}. The specific choices for the other parameters described in equations~\eqref{sub_m_f} and \eqref{sub_m_f_F} are outlined in Table \ref{params}.

We populate line-of-sight dark matter halos using the Sheth-Tormen (ST) mass function \citep{Sheth_2001}, incorporating two adjustments proposed by \cite{Gilman:2019vca}. The first adjustment, represented by a re-scaling factor $\delta_{\rm LOS}$, accommodates systematic fluctuations in the mean number of halos predicted by the Sheth-Tormen mass function and considers baryonic effects on small-scale structure formation. The second factor denoted as $\xi_{\rm 2halo}$, accounts for the two-halo term, which signifies the impact of correlated structures surrounding the host dark matter halo on the three-dimensional two-point correlation function \citep{Gilman:2019vca, gilman2021strong, Lazar_2021}. Consequently, the line-of-sight halo mass functions are expressed as
\be \label{Eq. 14}
   \frac{{\rm d}^2N_{\rm LOS}}{{\rm d}m{\rm d}V} = \de_{\rm LOS}\left(1+\xi_{\rm 2halo}(r, M_{\rm halo}, z) \right)\left[\frac{{\rm d}^2N_{\rm}}{{\rm d}m{\rm d}V}\right]_{\rm ST},  
\ee where
\be 
\xi_{\rm 2halo}(r, M_{\rm halo}, z) = b(M_{\rm halo}, z)\, \xi_{\rm lin}(r, z).
\ee The term $\xi_{\rm 2halo}(r, M_{\rm halo}, z)$ is determined by the halo bias $b(M_{\rm halo}, z)$ around the main halo as computed in \cite{Sheth_torman_1999}, and the linear matter correlation function $\xi_{\rm lin}(r, z)$ at a distance $r$, calculated using the linear matter power spectrum at redshift $z$. In each realization, \textsc {pyhalo} introduces negative convergence sheets along the line-of-sight and within the main lens plane to subtract the mean expected convergence from halos at each plane. This numerical procedure ensures that the Universe's mean density remains at the critical density after halos are added, while also restoring the Einstein radius of the lens --- accounting for dark matter substructure --- to its value in the smooth model. Without this correction, the inclusion of substructure leads to an over-density compared to the expected matter density of the Universe, causing an artificial focusing of light rays \citep{2017JCAP...04..049B}. This, in turn, could allow the network to infer the amount of substructure based on changes in the Einstein radius. The specific choices for the aforementioned parameters used to populate line-of-sight halos are also detailed in Table \ref{params}.

\subsubsection{Source}

To incorporate the source light into our simulated strong lens systems, we utilize galaxy images obtained from the \textit{HST} Cosmic Evolution Survey (COSMOS) \citep{2007ApJS..172....1S, 2007ApJS..172..196K}, captured with the \textit{HST} Advanced Camera for Surveys (ACS) using the F814W filter. These images have been processed by \textsc {paltas} \citep{Wagner-Carena:2022mrn}, which employs a subsample of postage stamp images \citep{2012MNRAS.420.1518M, 2014ApJS..212....5M} and applies additional selection criteria to ensure the inclusion of well-resolved galaxies in the subsample. For a detailed discussion of the selection criteria and cut parameters, the reader is referred to \cite{Wagner-Carena:2022mrn}.

From the COSMOS catalog, a subset of 2262 galaxies meets the selection criteria after the selection cuts. 
Among these, 2163 are designated for generating the training dataset, while 99 are set aside for validation and testing purposes, ensuring that the network is tested on sources it has not encountered during training. To create strong lens images, we randomly select one of these 2,262 images and utilize \textsc {paltas} to linearly interpolate the image for use with \textsc {lenstronomy}. Additionally, we apply a random rotation, and the Cartesian source position coordinates are determined by $(x_{\rm source}, y_{\rm source}) = \big(r_{\rm source} \sin(\theta_{\rm source}), r_{\rm source} \cos(\theta_{\rm source})\big)$ where $r_{\rm source}\sim \sqrt{\mathcal{U}(r^2_{\rm source, min}, r^2_{\rm source, max})}$ is the randomly selected radial polar coordinate and $\theta_{\rm source} \sim \mathcal{U}(0, \pi)$ radians is the angular coordinate, with $r_{\rm source, min}=0$ arcsec and $r_{\rm source, max}=0.2$ arcsec. In addition, we specified 1.5 for the source redshift $z_{\rm source}$.


After finalizing the selections for the main lens, dark matter halos, and source within the strong lens systems, we move forward to generate training, validation, and test datasets that mimic observations achievable with \textit{HST} capabilities. These datasets are constructed on $(80 \times 80)\ \text{pixel}^2$ grids, each corresponding to a physical dimension of  $(4 \times 4)\ \text{arcsec}^2$ and a resolution of $\rm 0.05 \, arcsec/pixel$. For the image modeling, we adopt the \textit{HST} ACS/WFC F814W filter configuration with an AB zero-point of 25.947. An empirical point spread function (PSF) with a full width at half maximum (FWHM) of 0.09 arcsec is applied to replicate the instrumental blurring. Additionally, uncorrelated noise is introduced across the pixel grid, including Poisson (shot) noise proportional to the photon count, considering an exposure time of 1428.0 seconds, and Gaussian background noise with an rms value of $\rm 0.006 \,counts/s/pixel$ added to each image.

\subsubsection{Two-point correlation function}

For each strong lens system generated as discussed previously, we compute the effective multiplane convergence map, denoted as $\kappa_{\text{div}}$, by taking the divergence of the effective deflection field, denoted as $\bm \alpha_{\rm eff}$ \citep{Gilman:2019vca, Dhanasingham:2023thg, Dhanasingham:2022nox}. This can be expressed as
\be \label{Eq. kappa div}
\kappa_{\rm div} \equiv \frac{1}{2} \bna \cdot \bal_{\rm eff} - \kappa_0. 
\ee To eliminate the lensing contribution from the main lens model, denoted as $\kappa_0$, we subtract the projected mass density of the single-plane main lens model from the divergence of the effective deflection field. This process removes the dominant contribution of the macrolens while preserving the non-linear coupling between the macrolens and line-of-sight halos. As discussed in \cite{Dhanasingham:2022nox} and \cite{Dhanasingham:2023thg}, line-of-sight dark matter halos in the strong lensed systems appear stretched in the tangential direction in the $\kappa_{\text{div}}$ maps due to the non-linear coupling between different lens planes in the system, thus exhibiting anisotropic signatures.

To quantify the anisotropies within the convergence map $ \kappa_{\text{div}} $ for the strong lens system using two-point correlation function multipoles $(\xi_{\ell})$, we adopt the approach delineated in the methodologies of \cite{Dhanasingham:2022nox} and \cite{Dhanasingham:2023thg}. Subsequently, we fit the monopole ($\ell=0$) and quadrupole ($\ell=2$) of the two-point function at small radial distances into a power law function represented by 
\be
\xi_\ell(r) = A_\ell \left(\frac{r}{r_0}\right)^{n_\ell} \hspace{5mm}{\rm for}\hspace{5mm}\ell=0,2,
\ee where $r_0 = (r_{\rm min}+r_{\rm max})/2$ serves as the pivot radius. We conduct the two-point function computation on convergence maps sized at $(250 \times 250)\ \text{pixel}^2$, utilizing a window size of 4 arcsec. For fitting the multipoles, the minimum radial distance $r_{\text{min}}$ considered is set to three times the resolution of the convergence map. This setting aims to eliminate wiggles occurring at small radial distances of correlation multipoles, as these wiggles arise when crossing the size of a pixel (resolution of the map), owing to the dominant numerical errors in computations at such small scales. Additionally, the maximum radial distance $r_{\text{max}}$ is chosen as six times the resolution of the convergence map to prevent errors associated with fitting multipoles into a linear function. After fitting, the statistics of the two-point function multipoles are thus described by four parameters: amplitudes $A_0$ and $A_2$, and slopes $n_0$ and $n_2$.




\section{Simulation-based inference (SBI)} \label{SBI}

To infer the posteriors of parameters of interest (learning parameters) from strong lens observations --- including the six main lens parameters ($\theta_{\rm E}, \, \gamma_{\rm macro}, $ main lens ellipticity components $e_1$ and $e_2$, and external shear Cartesian components $\, \gamma_1$ and $\gamma_2$), three parameters describing the properties of dark matter halos ($\Sigma_{\rm sub}, \, \delta_{\rm LOS},$ and $\beta$), and four parameters characterizing the statistics of the two-point function multipoles ($A_0, \, n_0, \, A_2$, and $n_2$) --- we employ the simulation-based inference (SBI) approach using a trained neural network as a density estimator, as discussed by \cite{Wagner-Carena:2022mrn}. SBI methods have gained popularity in parameter inference in cosmology due to their ability to take advantage of the full information content of complex data \citep{2015arXiv150602169C, Perreault_Levasseur_2017, Coogan:2020yux, Wagner-Carena_2021, Zhang:2022djp, Lemos:2022kua, 2023arXiv230915071M, Tucci:2023bag, 2023MNRAS.524.6167L, deSanti:2023rsw, Zhang:2023wda, Wagner-Carena:2024axc}. Following the procedure outlined in Section~\ref{data_gen}, we generate 400,000 mock observations for training, 1,000 validation data to ensure an unbiased evaluation of the neural density estimator's performance and to fine-tune its parameters during training, and an additional 1,000 images for testing the network's performance. Once trained, the network predicts the posterior distributions of thirteen key parameters, particularly those governing the structure and distribution of dark matter halos and the statistical properties of two-point function multipoles, based on a given strong lens image.

During training our neural density estimator, we minimize the loss function represented by 
\be
L(\phi) = -\sum_{k=1}^M \log q_{F(d_k, \phi)}(\xi_k)
\ee to evaluate and diagnose model optimization. Here, $ q_{F(d_k,\phi)}(\xi_k) $ represents the neural posterior density estimator, which is constructed based on the model architecture $ F(d_k, \phi) $ with network weights $ \phi $. It is trained using the image $ d_k $ and the associated parameters of interest $ \xi_k $ used to generate the image $ d_k $. Once trained, given a strong lens image $ d $ and a prior distribution $ \Omega_{\rm int} $ on the parameters of interest $ \xi $, the neural density estimator $ q_{F(d,\phi)}(\xi) $ predicts the posterior distribution of parameters $ p(\xi|d, \Omega_{\rm int}) $. Even though the prior distribution of learning parameters is not explicitly included in the loss function, it is indirectly learned through the distribution of $ \xi_k $ used to generate the training and validation images $ d_k $ \citep{Wagner-Carena:2022mrn}.

\subsection{Model architecture and training details}

In this work, we employ the \textsc {paltas} pipeline, which integrates the xResNet-34 architecture proposed by \cite{2015arXiv151203385H, 2018arXiv181201187H} and implemented using the \textsc {python} module \textsc {TensorFlow} \citep{2016arXiv160304467A}. The training process involves using a multivariate Gaussian as the density function $q$, with a full precision matrix of the Gaussian obtained through log-Cholesky decomposition. For a given number $N$ of parameters of interest, the final output layer of the xResNet-34 architecture is adapted to predict $N(N+3)/2$ parameters, comprising $N$ means of the Gaussian distributions of parameters of interest, $N$ outputs corresponding to the diagonal entries of the precision matrix, and $N(N-1)/2$ outputs corresponding to the off-diagonal entries of the precision matrix. More details on the methodology can be found in \cite{Wagner-Carena:2022mrn}. 

To facilitate architecture convergence, we preprocess the training images by normalizing them to have a standard deviation of 1. Additionally, we normalize the learning parameters to have a mean of zero and a unit standard deviation across the entire training dataset, saving these values to denormalize the predictions for the test data using the trained model. As highlighted by \cite{Wagner-Carena:2022mrn}, incorporating random rotation on each batch of images and subsequently adjusting the lensing parameters through the \textsc {paltas} pipeline enhances the quality of inference. Training our model involves dividing the training data into batches of 256 images each and employing Nvidia Tesla K40M dual GPUs. To estimate the mean and lower triangular elements of the precision matrix for $N=13$ parameters of interest, we train our neural posterior density estimator for 200 epochs over a total duration of approximately 56 hours. We employ the Adam optimizer with a learning rate of $10^{-5}$ and terminate the training process once the validation loss stabilizes below a threshold of $10^{-13}$.


%

\section{Results} \label{Results_4}

To assess the performance of our trained neural posterior density estimator in estimating the main lens parameters, as well as those describing the properties of dark matter substructure and the two-point function multipoles in individual lenses, we generate an additional set of 1,000 lenses using the parameters established for generating training and validation data, as outlined in Section~\ref{data_gen}. Subsequently, we employ our trained neural network to directly derive the posterior distributions of the parameters of interest from these test lenses. 
 
\subsection{Predicting substructure distribution and concentrations}

 \begin{figure}
\begin{center}
	\includegraphics[clip, trim=0cm 0cm 0cm 0cm, width=0.49\textwidth]{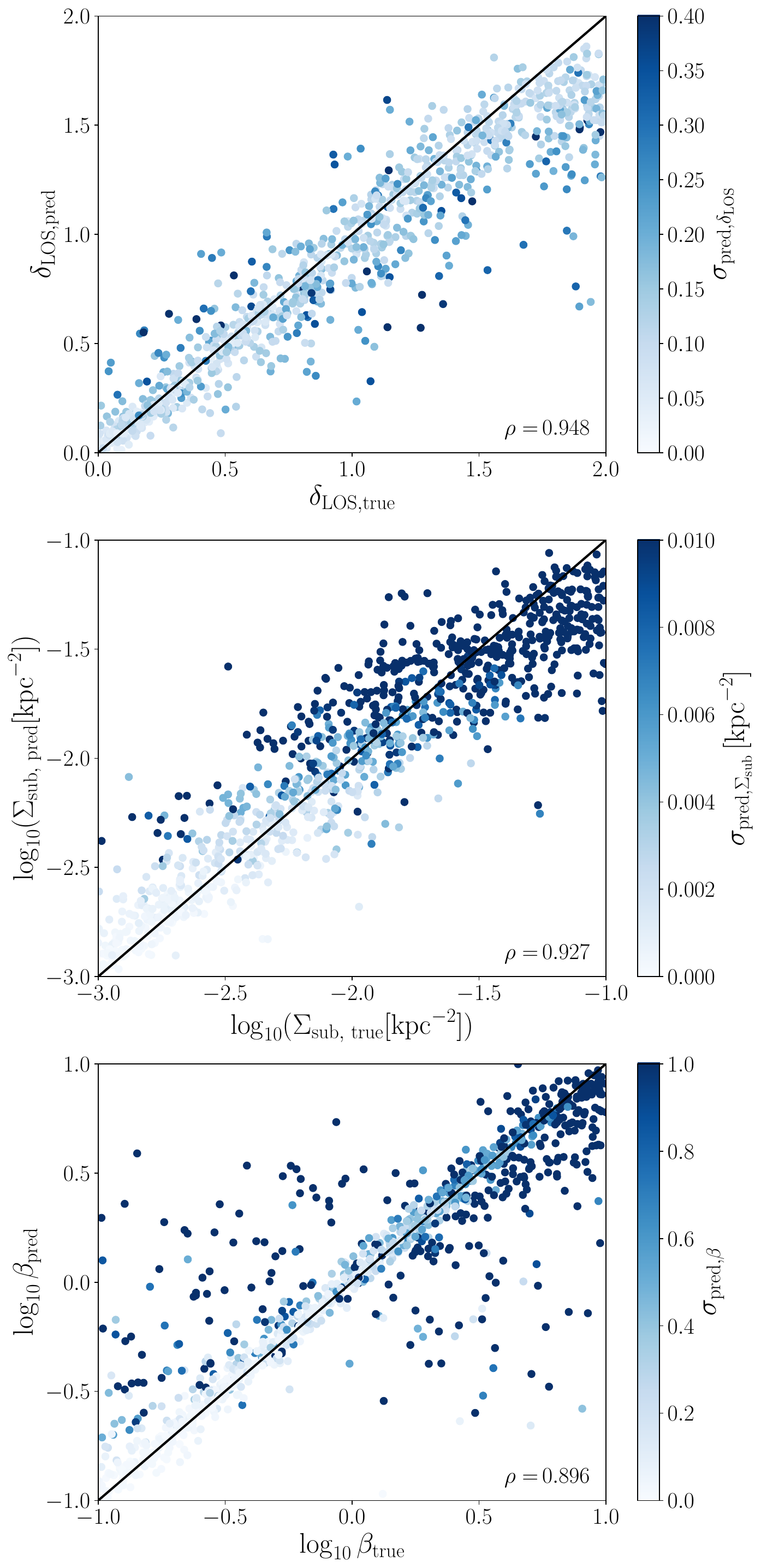}
\end{center}	
\caption{\label{fig:dm_params_true_pred} Comparison of the predicted and true mean values for three parameters describing dark matter halo properties ($\delta_{\rm LOS}, \,\Sigma_{\rm sub},$ and $\beta$) across 1,000 test lenses. Each point represents a single lens, with the Pearson correlation coefficient ($\rho$) displayed in the bottom-right corner of each plot, quantifying the correlation between the neural density estimator’s predictions and the true values. The color of each point corresponds to the predicted standard deviation. Black lines indicate where the predicted values align with the true values.}
\end{figure}

\begin{figure*}
\begin{center}
	\includegraphics[clip, trim=0cm 0cm 0cm 0cm, width=0.95\textwidth]{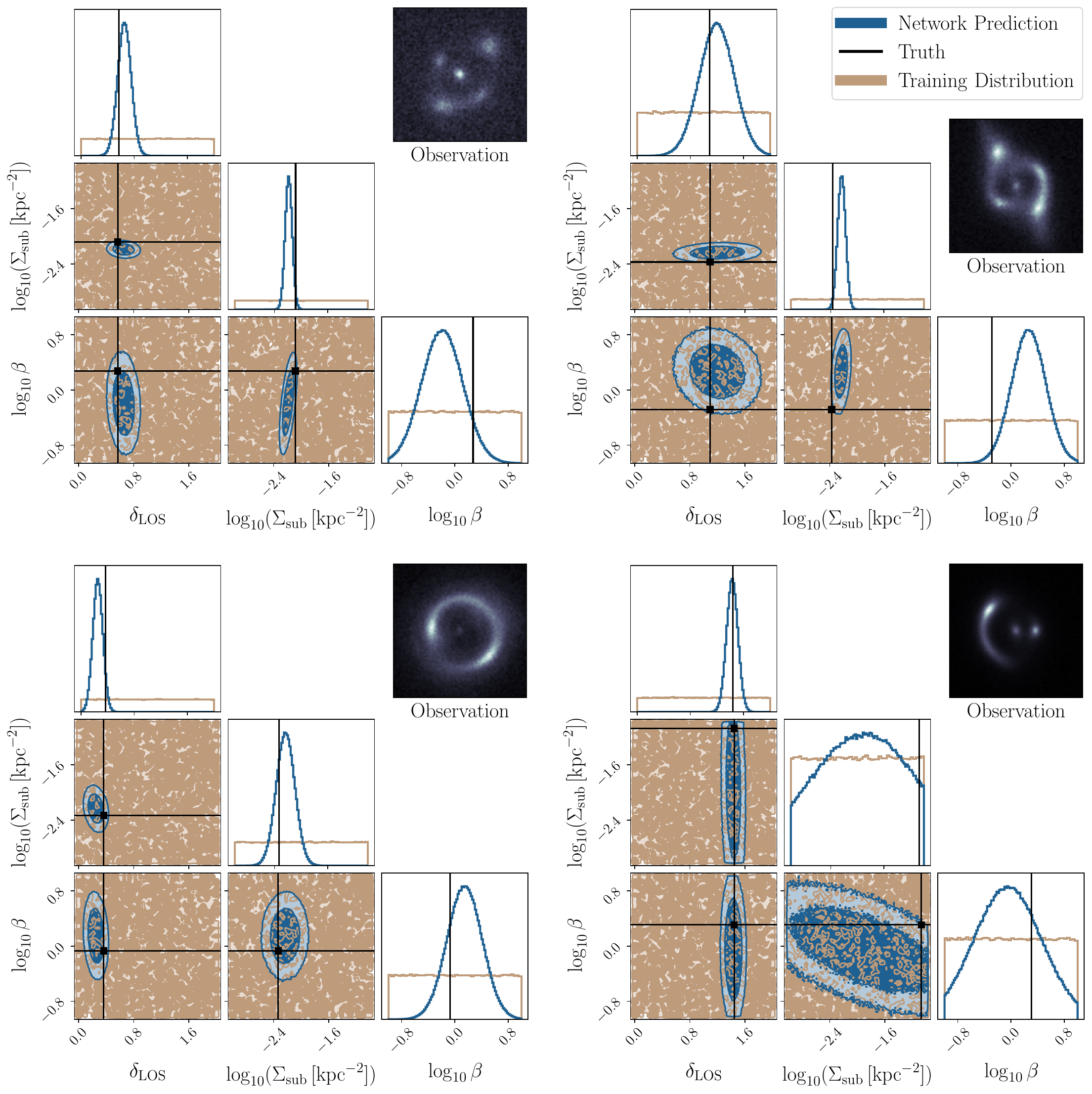}
\end{center}	
\caption{\label{fig:dm_params_corner_reandom}Posterior distributions estimated by the trained neural network for parameters describing dark matter substructure from four different test images, shown in the upper-right corner of each panel. This figure illustrates the neural posterior density estimator's ability to predict substructure abundances and concentrations under different conditions of low/high subhalo and line-of-sight halo abundances, as well as low/high halo concentrations. Blue contours represent the neural network's predictions, while beige contours indicate the prior distributions used in the training data. Dark and light contours correspond to the 68\% and 95\% confidence intervals, respectively. The true parameter values used to generate the strong lens images are marked by black points.}
\end{figure*}

Fig.~\ref{fig:dm_params_true_pred} illustrates the correlation between the true and predicted values derived from our trained neural posterior density estimator for the line-of-sight mass function re-scaling parameter ($\delta_{\rm LOS}$; top panel), the subhalo mass function normalization ($\Sigma_{\rm sub}$; middle panel), and the free overall scaling factor of the Diemer and Joyce mass–concentration relationship ($\beta$; bottom panel). Each point corresponds to an individual test lens, with the color indicating the predicted standard deviation ($\sigma_{\rm pred}$) of the Gaussian output by the neural posterior density estimator. Because the network outputs for both $\Sigma_{\rm sub}$ and $\beta$ (denoted collectively as $\Omega$) follow log-normal distributions, the color scale represents the standard deviation converted to the linear domain ($\sigma_{\rm pred, \Omega}$), computed as $\sqrt{[\exp(\sigma_{\rm pred, \ln \Omega}^2)-1]\exp(2\mu_{\rm pred, \ln \Omega}+\sigma_{\rm pred, \ln \Omega}^2)}$, where $\mu_{\rm pred, \ln \Omega}$ denotes the predicted mean of the Gaussian output by the network. The Pearson correlation coefficient ($\rho$) is displayed in the bottom-right corner of each panel. Overall, the predicted mean values for all three parameters are well-constrained, as indicated by high Pearson correlation coefficients. However, among the three, $\beta$ exhibits comparatively weaker constraints, characterized by a higher degree of scatter. Fig.~\ref{fig:dm_params_corner_reandom} presents the predicted posterior distributions for these parameters in four randomly selected lenses from the test dataset. These lenses exhibit varying subhalo and line-of-sight halo abundances and concentrations, which directly influence the magnitude of subtle perturbations in the lensed images and, consequently, the network's ability to infer these parameters. Overall, the results demonstrate that our trained neural posterior density estimator yields reasonably good results, effectively inferring the parameters. The posterior distribution is not overly influenced by the training data distribution, allowing maximum information extraction from each individual lens. However, the challenges arise from model misspecification and the complexities of real data still impact the performance of the SBI approaches. Since these simulated strong lens systems are approximations, they may not exactly replicate real data, and any discrepancies between the simulated and real data can lead to biased or overly confident posterior estimates \citep{2025A&A...694A.212B, 2025arXiv250923385R}.

The network learns by systematically analyzing strongly lensed images within the training dataset, where the subtle, localized gravitational perturbations---originating from substructure beyond the smooth mass distribution of the main lens galaxy---play a critical role in shaping the learning outcomes. These perturbations are particularly sensitive to the projected mass density of substructures near the Einstein radius, the region where lensed images form. Although higher values of $\delta_{\rm LOS}$, $\Sigma_{\rm sub}$, and $\beta$ generally correspond to stronger perturbations and thus yield tighter constraints on these parameters, the results presented in the middle and bottom panels of Fig.~\ref{fig:dm_params_true_pred} indicate that the network achieves better recovery of $\Sigma_{\rm sub}$ and $\beta$ with smaller predicted standard deviations $\sigma_{\rm pred, \Sigma_{\rm sub}}$ and $\sigma_{\rm pred, \beta}$ for test lens systems characterized by lower $\Sigma_{\rm sub}$ and $\beta$ values. This is somewhat counterintuitive, as one might expect systems with a greater number of subhalos and more concentrated halos to be better constrained.
Indeed, as expected, both the over-predicted and the under-predicted $\beta$ exhibit higher predicted standard deviations, reflecting greater uncertainty in these cases.

\begin{figure}
\begin{center}
	\includegraphics[clip, trim=0cm 0cm 0cm 0cm, width=0.49\textwidth]{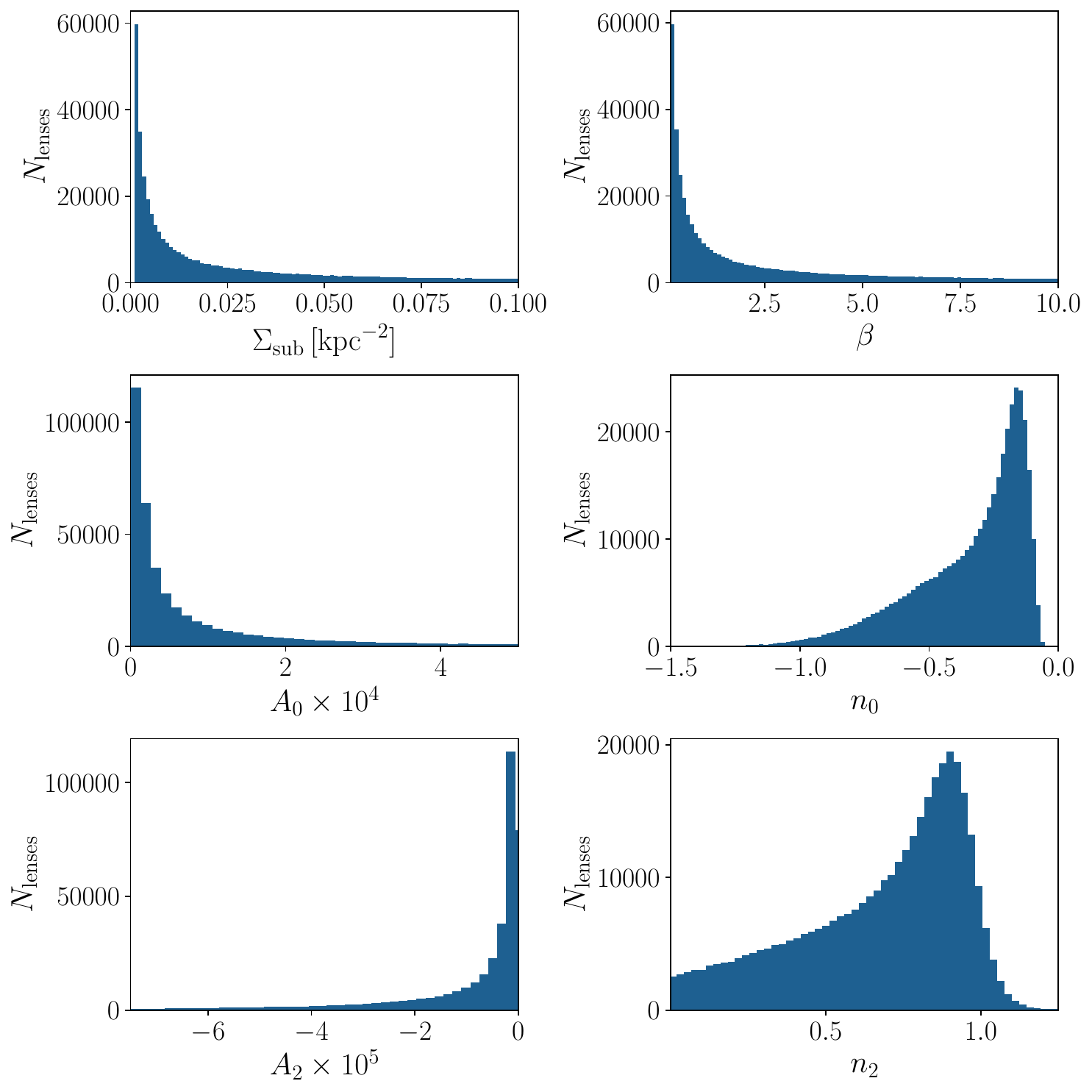}
\end{center}	
\caption{\label{fig:Tr_hist} Histograms illustrating the distributions of the parameters $\Sigma_{\rm sub}$, $\beta$, $A_0$, $n_0$, $A_2$, and $n_2$ within the training dataset. The pronounced skewness in these distributions may introduce biases in the neural posterior density estimator, potentially leading it to favor outputs that mirror the underlying training set distribution.}
\end{figure}

This apparent discrepancy is attributable to the relative scarcity of high $\Sigma_{\rm sub}$ and $\beta$ systems within the training set. This imbalance arises because $\Sigma_{\rm sub}$ and $\beta$ are sampled from log-uniform distributions, resulting in a heavily skewed data distribution when transformed into the linear domain (see Fig.~\ref{fig:Tr_hist}). Consequently, this bias in the training data limits the performance of the network in parameter regimes with large $\Sigma_{\rm sub}$ and $\beta$ values, as reflected by the larger predicted uncertainties. This highlights how the choice of training distribution can significantly affect predictive accuracy across different parameter regimes, as previously detailed by \cite{2021A&A...646A.126S}, examining how training distributions centered around $\theta_{\rm E}$ impact model performance. Although the presence of a substructure reliably induces localized perturbations in the lensed images, these signals often remain subdominant compared to observational noise. Moreover, even when subhalo and line-of-sight halo mass functions are fixed, the amplitude of these perturbations strongly depends on the projected number density of halos overlapping the lensed images. These factors collectively contribute to the considerable scatter observed in Fig.~\ref{fig:dm_params_true_pred}. Additionally, since $\delta_{\rm LOS}$ in the training data was sampled from a uniform distribution, the predicted standard deviations ($\sigma_{\rm pred, \delta_{\rm LOS}}$) for test lenses do not exhibit a clear trend, unlike the behavior observed for $\sigma_{\rm pred, \Sigma_{\rm sub}}$ and $\sigma_{\rm pred, \beta}$.

\begin{figure*}
\begin{center}
	\includegraphics[clip, trim=0cm 0cm 0cm 0cm, width=0.85\textwidth]{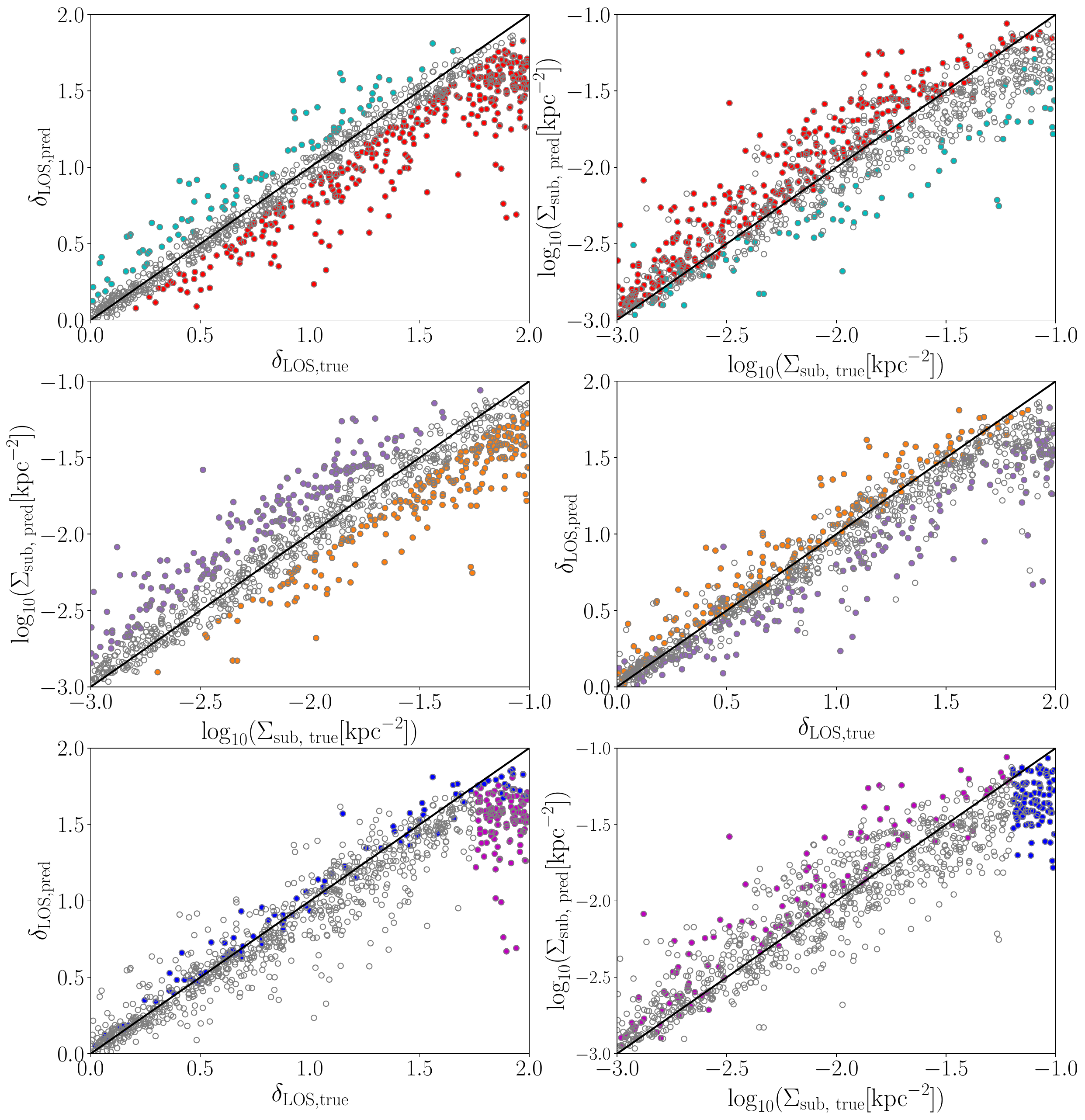}
\end{center}	
\caption{\label{fig:DM_above_below} In the top and middle rows, the left panels highlight points with overpredictions and underpredictions using color coding, while the right panels show the corresponding locations of these points in the other parameter space. These panels illustrate an inverse relationship between the predicted line-of-sight dark matter contribution ($\delta_{\rm LOS}$) and the subhalo mass function normalization ($\Sigma_{\rm sub}$): underestimating one tends to coincide with overestimating the other. At higher values, both parameters deviate from the one-to-one relation, indicating systematic underpredictions. In the bottom row, points that show significant deviations are color-coded, and their locations are displayed in the complementary parameter space.This highlights the consistency of the trend across lens systems and further supports the degeneracy between $\delta_{\rm LOS}$ and $\Sigma_{\rm sub}$.}
\end{figure*}

As shown in the top and middle panels of Fig.~\ref{fig:DM_above_below}, strong lensing systems with underpredicted $\delta_{\rm LOS}$ lead to an overprediction of the subhalo mass function normalization, $\Sigma_{\rm sub}$, while overpredicted $\delta_{\rm LOS}$ results in an under-prediction of $\Sigma_{\rm sub}$, and vice versa. Additionally, at larger values, the predicted mass function parameters ($\Sigma_{\rm sub}$ and $\delta_{\rm LOS}$) deviate noticeably from the reference line, indicating an underprediction in this regime. The lower panel of Fig.~\ref{fig:DM_above_below} further illustrates that lens systems that show this behavior with larger $\delta_{\rm LOS}$ exhibiting underpredictions correspond to an overprediction of $\Sigma_{\rm sub}$, and vice versa. Although line-of-sight dark matter halos introduce distinctive anisotropic signatures in strong lensing systems, these plots reveal a degeneracy between $\delta_{\rm LOS}$ and $\Sigma_{\rm sub}$ in our test dataset, where decreasing one parameter while increasing the other produces similar observational effects. Finally, we find no significant degeneracy between the concentration parameter $\beta$ and either the parameters $\delta_{\rm LOS}$ and $\Sigma_{\rm sub}$.

\begin{figure*}
\begin{center}
	\includegraphics[clip, trim=0cm 0cm 0cm 0cm, width=1\textwidth]{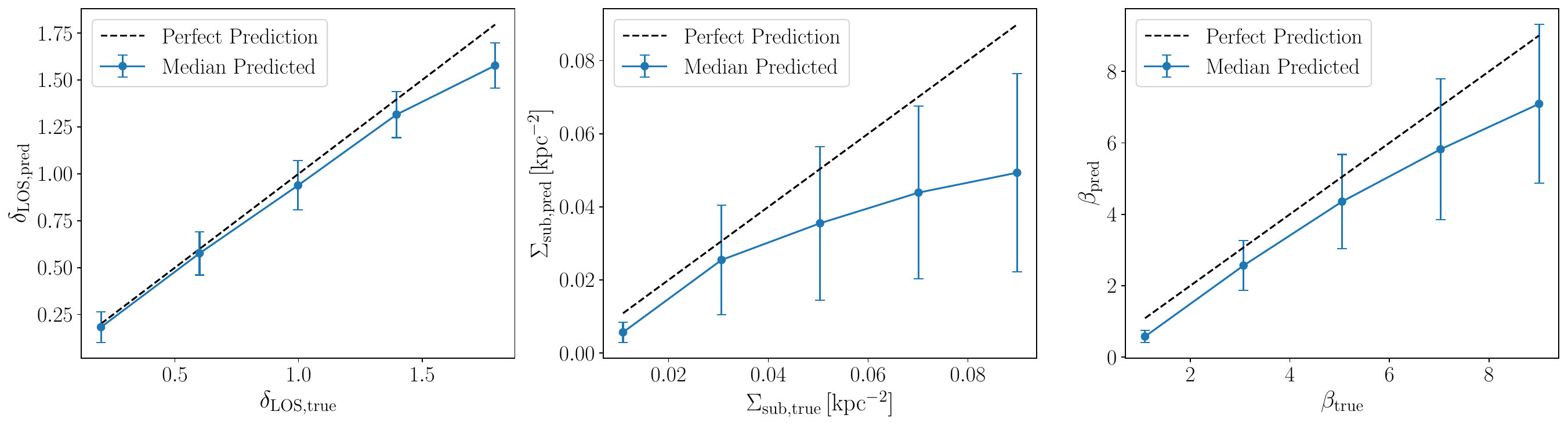}
\end{center}	
\caption{\label{fig:True_vs_pred_binned} Binned median-per-bin predicted vs. true values for parameters $\delta_{\rm LOS}$, $\Sigma_{\rm sub}$, and $\beta$. Each point represents the median of the predicted mean values, with an error bar showing the median of the predicted standard deviations in each bin, plotted against the bin center of the true values. The left-skewed distributions of $\Sigma_{\rm sub}$ and $\beta$ in the training dataset lead to systematic underprediction for large values, indicating a bias induced by the data distribution. $\delta_{\rm LOS}$, although uniformly distributed in the training set, also exhibits bias at high values due to degeneracy with $\Sigma_{\rm sub}$.}
\end{figure*}

In addition to exhibiting large predicted standard deviations for high values of $\Sigma_{\rm sub}$ and $\beta$, the network’s predicted means show a systematic trend of under-prediction for these parameters, despite the expectation that the network could more accurately recover them, particularly for large values, since these parameters can generate strong perturbations in the lensed images. In Fig.~\ref{fig:True_vs_pred_binned}, the true values of $\delta_{\rm LOS}$, $\Sigma_{\rm sub}$, and $\beta$ in the test set are divided into five linearly spaced bins. For each bin, the median of the predicted mean values is plotted against the bin center of the corresponding true values, with error bars indicating the median of the predicted standard deviations. The network’s reduced performance for large $\Sigma_{\rm sub}$ and $\beta$ can be attributed to the highly skewed distributions of these parameters in the training dataset, which contains a larger fraction of lenses with small values. Consequently, the network is biased toward predicting smaller $\Sigma_{\rm sub}$ and $\beta$ more frequently, while large values are underrepresented. A similar behavior has been reported by \cite{Wagner-Carena:2022mrn}, where the network tends to systematically over- or under-predict $\Sigma_{\rm sub}$, resulting in predictions that cluster around the mean of the training distribution of $\Sigma_{\rm sub}$, regardless of the true values in the test data. Additionally, the degeneracy between $\Sigma_{\rm sub}$ and $\delta_{\rm LOS}$ contributes to the observed under-prediction of $\Sigma_{\rm sub}$ at larger true values, as shown in Fig.~\ref{fig:True_vs_pred_binned}, and similarly explains the under-prediction of $\delta_{\rm LOS}$ for lenses with larger true values.




\subsection{Predicting two-point function multipoles}

In this section, we present the results of using the trained neural posterior density estimator, applied in the previous section to infer the hyperparameters of dark matter substructure, to infer the summary statistics that characterize the two-point multipoles, which are crucially independent of the specific parameterization of the dark matter model.

\begin{figure*}
\begin{center}
	\includegraphics[clip, trim=0cm 0cm 0cm 0cm, width=0.75\textwidth]{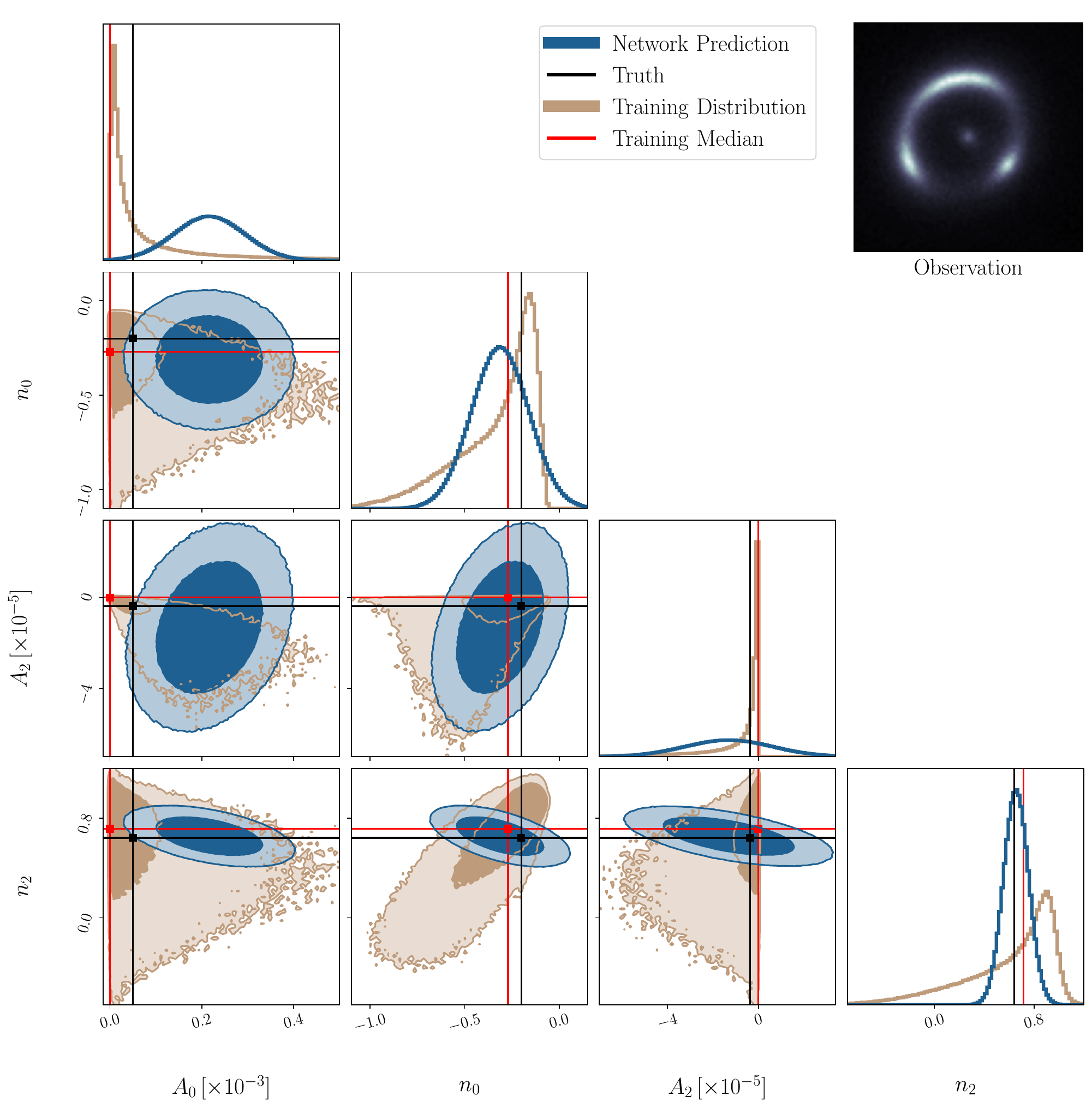}
\end{center}	
\caption{\label{fig:corner} This figure illustrates the neural posterior density estimator's output for the parameters characterizing the two-point function multipoles for a simulated lens image from our testing dataset shown in the upper-right corner. The blue contours represent the network's predictions, while the beige contours depict the prior distribution of the parameters used in the training data. The dark and light contours signify the 68\% and 95\% confidence intervals, respectively. True values used to generate the strong lens image are denoted by black points.} 
\end{figure*}

In Fig.~\ref{fig:corner}, we observe that while the neural density estimator is capable of constraining the parameters describing the distribution and concentration of dark matter substructure --- even in the presence of challenges such as training data distribution bias and the degeneracy between $\delta_{\rm LOS}$ and $\Sigma_{\rm sub}$ --- the trained network also exhibits some sensitivity to the two-point correlation function parameters ($A_0$, $n_0$, $A_2$, and $n_2$), albeit with greater associated uncertainties. Notably, the predicted multivariate Gaussian posteriors are not fully dominated by the parameter distribution in the training data. The predicted means of these parameters do not merely cluster around the medians of their training distributions, which indicates that the network is not just learning the priors, but is genuinely extracting signal from the data. This behavior highlights the network’s capacity to extract meaningful information from individual lensing systems, even for parameters that are more weakly constrained.

\begin{figure*}
\centering
\includegraphics[width=0.95\textwidth]{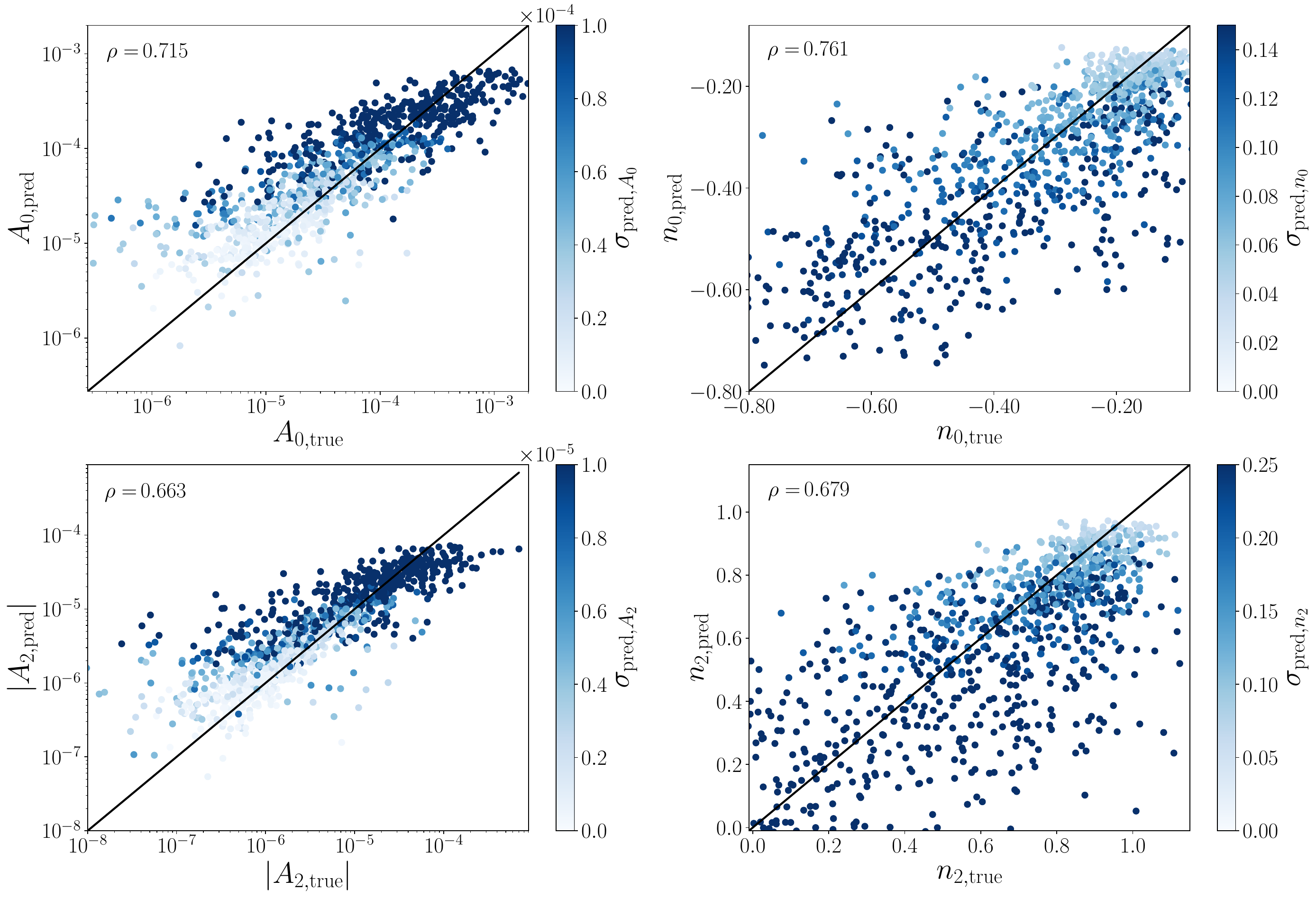}
\caption{\label{fig:corr_stat_true_pred} Similar to Fig.~\ref{fig:dm_params_true_pred}, this figure presents a comparison between the predicted mean values of parameters describing two-point function multipoles and their true values for 1000 test lenses. The Pearson correlation coefficient, $\rho$, shown in the top-left corner of each plot, quantifies the correlation between the neural density estimator's predictions and the actual values.}
\end{figure*}

Fig.~\ref{fig:corr_stat_true_pred} presents a comparison between the predicted mean values and the true values of the two-point function multipole parameters across 1,000 test lenses. Although all four parameters exhibit moderately strong correlations with their ground-truth values, the predictions display considerable scatter, making it difficult to constrain these parameters precisely. This behavior is consistent with the trends observed in Fig.~\ref{fig:dm_params_true_pred}, where the isotropic lensing signatures—arising from both subhalos and line-of-sight halos and encoded by $A_0$ and $n_0$ --- and the anisotropic contributions --- due to line-of-sight halos and captured by $A_2$ and $n_2$ --- are generally subdominant relative to observational noise. Furthermore, the ability to extract these features depends more critically on the projected number density of halos that overlap with the lensed images than on the broader annular region used to compute the true parameter values. In addition, line-of-sight halos that lie near the main lens plane typically generate weak anisotropic signatures that are often absorbed into the monopole terms. As a result, the quadrupole parameters tend to show weaker correlations and thus lower Pearson correlation coefficients compared to their monopole counterparts.

Crucially, the distributions of the two-point function parameters in the training dataset are not arbitrarily selected, but instead are derived from the projected mass density ($\kappa_{\rm div}$) maps of the lensed systems. These distributions are largely influenced by the priors adopted for the substructure mass function and concentration parameters. Although we retain some control over the resulting parameter distributions through our choice of priors, particularly the normalization of the substructure mass functions and the re-scaling of halo concentrations, we cannot exert complete control. In cases where the training data are skewed toward certain parameter values, the network tends to predict narrower posteriors for test lenses whose true values lie near these regions of skewness.

\begin{figure*}
\begin{center}
	\includegraphics[clip, trim=0.2cm 0cm 0cm 0cm, width=0.9\textwidth]{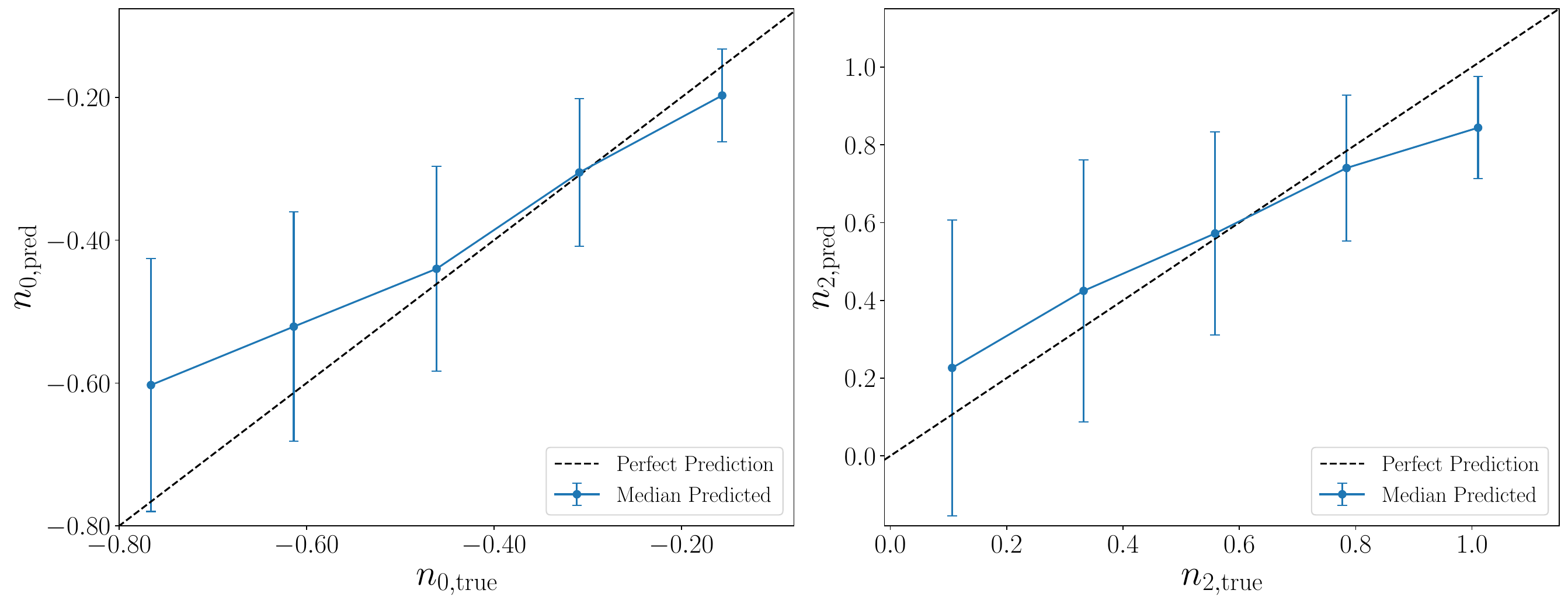}
\end{center}	
\caption{\label{fig:True_vs_pred_binned_2pcf} Similar to Fig.~\ref{fig:True_vs_pred_binned}, but showing the power-law slope parameters $n_0$ and $n_2$. The skewed distributions of these parameters in the training dataset lead to systematic biases in the predictions.}
\end{figure*}

Consistent with the biases observed in the predicted means and the standrad deviations of the dark matter substructure parameters, the network's predictions for $A_0$, $n_0$, $A_2$, and $n_2$ also exhibit systematic trends of under- or overestimation, depending on the underlying distribution of the training data. The network consistently underpredicts the multipole amplitudes, particularly for lenses with large true absolute values of $A_0$ and $A_2$, reflecting the training dataset's bias toward lower absolute values. Figure~\ref{fig:True_vs_pred_binned_2pcf} shows the binned medians of the predicted values of $n_\ell$ plotted against the corresponding bin centers of the true values in the test set. As illustrated in the figure, the power-law slope parameters $n_0$ and $n_2$ tend to be either under- or overestimated, depending on how their true values compare with the more populated regions of the training distribution. These prediction biases, compounded by observational noise, present a significant challenge to accurately inferring the two-point correlation function parameters using our trained neural posterior density estimator. 




In general, lens systems with very few line-of-sight halos typically produce only weak anisotropic signatures, so both the quadrupole amplitude $A_2$ and slope $n_2$ are correspondingly small. Fitting a power law in the regime $n_2 \to 0$ is numerically unstable, often leading to unreliable or high-variance estimates of $n_2$. Consequently, while the network may predict values ($n_{2, \rm pred}$) that broadly agree with the true ones ($n_{2, \rm true}$) used for comparison, these $n_{2, \rm true}$ and $n_{2, \rm pred}$ can still deviate significantly from the quadrupole measured directly from the $\kappa_{\rm div}$ maps and, in fact, do not provide an accurate representation of this actual quadrupole.



 
\subsection{Revisiting macrolens parameter estimation using SBI}

As a final remark, the trained network effectively constrains the main lens parameters in the test dataset, with the predicted mean values exhibiting a strong correlation with the true values. These results align well with previous work on this topic \citep[see e.g.,][]{Perreault_Levasseur_2017, 2019MNRAS.488..991P, 2021ApJ...910...39P, 2021MNRAS.505.4362P, Wagner-Carena_2021, Wagner-Carena:2022mrn}. Appendix~\ref{App:A} provides further illustration of these results: Fig.~\ref{fig:corner_macro} presents the predicted posterior distributions for the main lens parameters of a randomly selected lens from the test dataset, while Fig.~\ref{fig:main_lens_true_pred} compares the true values with the network-predicted means of the macrolens parameters. 

Among the key main lens parameters, $\gamma_{\rm macro}$ shows a relatively weaker correlation with the true values in our test dataset ($\rho = 0.782$), as shown in Fig. \ref{fig:main_lens_true_pred}. Additionally, the correlation between the true and predicted values for $\gamma_1$ and $\gamma_2$ is also comparatively weaker ($\rho \sim 0.88$). This can mainly be attributed to the impact of the inclusion of the main lens light during both the network training and testing. \cite{2019MNRAS.488..991P} demonstrated a 34\% average improvement in accuracy when predicting the macrolens parameters after removing lens light in their non-Bayesian CNN approach. This is particularly true when the CNN is trained with the main lens light included, as the model becomes influenced by the lens rather than the source light, causing it to learn from the main lens light, rather than from the lensed images. Even when lens light is removed, \cite{2021ApJ...910...39P} observed that partial mixing of lensed source images with main lens light still negatively affects the prediction accuracy of their trained Bayesian Neural Network (BNN), especially for $\gamma_{\rm macro}$. As \cite{2019MNRAS.488..991P} noted, this performance drop due to lens light inclusion can be mitigated by incorporating multi-band imaging. They further showed that introducing a moderate amount of misalignment between the lens light and mass profile ellipticities enhances the robustness of predictions for ellipticity, preventing the network from being more influenced by the lens light rather than the lensed source light. In line with this, we find that sampling the lens and mass profile axis ratios and orientations from two independent distributions allows us to constrain $e_1$ and $e_2$ more effectively.

In addition to the effects of including lens light, the diminished correlation observed when recovering these main lens parameters could potentially be caused by the mass-sheet degeneracy (MSD), which arises from the source position transformation \citep{1995A&A...294..411S, 2014A&A...564A.103S, 2017A&A...601A..77U, 2018A&A...617A.140W}, where the simultaneous transformation of both the source and lens planes can leave several astrometric and photometric lensing observables invariant. MSD makes it inherently difficult to precisely measure $\gamma_{\rm macro}$ \citep[see e.g.][]{2013A&A...559A..37S, 2020JCAP...11..045G}, and it also rescales the external shear components as  $ \gamma \rightarrow \lambda \gamma $ under transformation of the coordinates in the source plane $\uu \rightarrow \lambda \uu$ \citep{{2016MNRAS.460.2505R}}. This potential source of bias, which has not yet been fully explored in the context of SBI approaches, may lead to the ability to explain a given test image using different values of the same parameter. For instance, two different combinations of image positions and $\gamma_{\rm macro}$ (or $\gamma$) could explain the same observation. Additionally, the degeneracy between $\gamma$ and the axis ratio ($q_e$) \citep{1997MNRAS.291..211W, 2024MNRAS.531.3684E} 
could further complicate the accurate estimation of $\gamma$, potentially biasing the network's predictions. As a result, these effects contribute to poorer recovery of the external shear and the power-law slope of the main lens compared to other macrolens parameters.

\section{Discussion and conclusions} \label{Discussion}

Galaxy-scale strong gravitational lenses provide sensitive probes of dark matter substructure, which introduce subtle perturbations to the lensed images. Analysing these perturbations offers a unique opportunity to investigate dark matter on smal cosmological scales. Previous work by \cite{Dhanasingham:2023thg, Dhanasingham:2022nox} highlighted that line-of-sight dark matter halos in strong lens systems exhibit characteristic anisotropic signatures in the projected mass density maps, as captured under the effective multiplane lensing framework. These signatures can be identified as a non-zero quadrupole moment in the two-point correlation function of the effective convergence maps. As demonstrated in these studies, both isotropic lensing perturbations from line-of-sight halos and subhalos, as well as anisotropic lensing effects from line-of-sight dark matter halos, are highly sensitive to dark matter self-interactions and substructure abundances. Consequently, the multipoles of the two-point function offer a unique and powerful method for studying dark matter microphysics. In this study, we explore the effectiveness of a trained convolutional neural network (CNN) in learning and analyzing these perturbations, enabling likelihood-free inference of the two-point correlation function multipoles derived from strong gravitational lens systems. To train the network, we generated extensive datasets comprising 400,000 simulated lenses for training and 1,000 for validation. Source galaxies were drawn directly from the COSMOS field to capture realistic morphological complexities, while observational and instrumental effects have been incorporated to ensure the fidelity of the simulated strong lens images, akin to those observed by the \textit{HST}.

To evaluate the trained neural posterior density estimator’s performance on individual lenses, we generated an additional set of 1,000 lenses. For each lens, the network predicted the dark matter substructure parameters --- namely the line-of-sight mass function re-scaling parameter ($\delta_{\rm LOS}$), the subhalo mass function normalization ($\Sigma_{\rm sub}$), and the free scaling factor of the Diemer and Joyce mass-concentration relationship ($\beta$) --- with reasonably good accuracy. The network effectively infers these parameters, with strong correlations observed between the true values and the network-predicted mean values of the Gaussian posterior distributions. However, both the predicted means and standard deviations of these parameters are significantly influenced by their distributions in the training dataset. Specifically, the skewed distributions of $\Sigma_{\rm sub}$ and $\beta$ in the training data, compared to $\delta_{\rm LOS}$, which is sampled from a uniform distribution, lead to smaller predicted standard deviations for test lenses near the mode of the distributions. In contrast, test lenses whose true parameter values lie in the tails of the skewed distributions are predicted less accurately. This indicates that the network tends to favor predictions near the modes of the distributions, while values outside this range are underrepresented. This underscores how the training data distribution can significantly impact the model’s accuracy across different parameter regimes. Additionally, a degeneracy is observed between $\delta_{\rm LOS}$ and $\Sigma_{\rm sub}$ in the test dataset, where underestimating one parameter often coincides with overestimating the other, suggesting that changes to one parameter, while adjusting the other, result in similar observational effects, even though line-of-sight dark matter halos typically produce distinct anisotropic signatures in strong lensing systems. This degeneracy, along with the other factors, makes predicting the parameters that describe the properties of dark matter substructure particularly challenging.

At the individual lens level, the trained network demonstrates some sensitivity to the two-point correlation function parameters ($A_0$, $n_0$, $A_2$, and $n_2$), although with higher associated uncertainties. The predicted means of these parameters do not simply cluster around the medians of their training distributions, indicating that the network is not merely learning the priors but is instead extracting meaningful signals from the data. This behavior underscores the network's ability to capture some information about dark matter substructure from individual lensing systems, even for parameters that are less tightly constrained. Importantly, the distributions of the parameters $A_0$, $n_0$, $A_2$, and $n_2$ in the training dataset are not arbitrarily chosen, but are derived from the $\kappa_{\rm div}$ maps of the lensed systems. As such, these distributions are largely shaped by the priors selected for the substructure mass function and concentration parameters. Although we can influence the resulting parameter distributions through our choice of priors --- particularly for $\delta_{\rm LOS}$, $\Sigma_{\rm sub}$, and $\beta$ --- we cannot exert complete control. Consistent with the biases observed in the predicted means and standard deviations of $\delta_{\rm LOS}$, $\Sigma_{\rm sub}$, and $\beta$, the network’s predictions for $A_0$, $n_0$, $A_2$, and $n_2$ also exhibit systematic trends of under- or over-estimation, depending on the underlying distribution of these parameters in the training data. Additionally, lens systems with minimal substructure typically produce weak perturbations, leading to smaller amplitudes and slopes in the two-point correlation multipoles. Fitting a power-law in the regime $n_\ell \to 0$ is numerically unstable and often results in unreliable or high-variance estimates of $n_\ell$. Therefore, while the network’s predicted values ($n_{\ell,\rm pred}$) may broadly agree with the true values ($n_{\ell,\rm true}$) used for comparison, these values can still deviate significantly from the multipoles directly measured from the $\kappa_{\rm div}$ maps. In fact, fitting these multipoles to a power-law function may not provide an accurate representation of the actual multipoles.

In addition to the effects of data distribution bias and the degeneracy between $\delta_{\rm LOS}$ and $\Sigma_{\rm sub}$ on accurately predicting dark matter substructure and two-point function parameters using galaxy-scale strong lenses, the perturbations induced by dark matter substructure often remain subdominant compared to observational noise, making it challenging to constrain these parameters precisely. Moreover, even when the subhalo and line-of-sight halo mass functions are fixed, the amplitude of these perturbations strongly depends on the projected number density of halos overlapping the lensed images. Additionally, line-of-sight halos near the main lens plane typically produce weak anisotropic signatures, which are often absorbed into the isotropic components. These factors result in significant scatter between the true and predicted values, further complicating precise parameter constraints. Furthermore, model misspecification and the inherent complexities of real data continue to pose significant challenges, impacting the performance of SBI approaches in studying dark matter substructure through galaxy-scale strong lensing observations. Since the simulated lens systems are approximations, they may not perfectly reflect real data, and discrepancies between the simulated and actual lenses can lead to biased or overly confident posterior estimates.

\cite{2019MNRAS.488..991P} reported a significant improvement in the accuracy of the recovery of main lens parameters when lens light was removed before training and testing their neural networks. This reduction in accuracy can be mitigated by incorporating color information, which allows the network to learn the color differences between the lens and the source in the images. We believe that the scatter observed in our network predictions of dark matter substructure and two-point function parameters is significantly influenced by the inclusion of lens light, as the network tends to learn these parameters from the main lens light rather than from the source light. Additionally, as discussed by \cite{Dhanasingham:2022nox}, the two-point correlation function is affected by the mass-sheet degeneracy (MSD), which biases the recovery of the two-point function. Therefore, addressing MSD and incorporating methods like independent component analysis \citep[ICA;][]{Hezaveh_2017} to remove lens light before passing the images to train the network, along with including multi-band imaging to mitigate the effects of lens light, seems like a promising avenue for future work.

As discussed, our network shows limitations in effectively constraining the two-point function statistics. As a result, we defer the exploration of alternative methods to more accurately characterize these multipoles for future investigation. Additionally, relying solely on a multivariate Gaussian posterior may be insufficient in capturing the full statistical complexity of the two-point function multipoles. Therefore, incorporating normalizing flows into this framework appears to be a promising direction for future research. As highlighted by \cite{Dhanasingham:2023thg}, given that the amplitudes and slopes of the two-point function multipoles are highly sensitive to dark matter self-interactions, combining a well-established likelihood-free inference method with a linearized likelihood-based approach could offer a robust strategy for constraining dark matter self-interactions using galaxy-scale strong lenses.

\section*{Acknowledgements}

We thank Annika H. G. Peter, Charlie Mace, and Bharath Chowdhary Nagam for useful discussions. B. D. and F.-Y. C.-R. acknowledge the support of program HST-AR-17061.001-A whose support was provided by the National Aeronautical and Space Administration (NASA) through a grant from the Space Telescope Science Institute, which is operated by the Association of Universities for Research in Astronomy, Incorporated, under NASA contract NAS5-26555. D. G. acknowledges support from the Brinson Foundation through a Brinson Prize Fellowship grant. F.-Y. C.-R. would like to thank the Robert E. Young Origins of the Universe Chair fund for its generous support. F.-Y. C.-R. also acknowledges the support of National Science Foundation award OIA-2327192. We also would like to thank the UNM Center for Advanced Research Computing (CARC), supported in part by the National Science Foundation, for providing the high-performance computing resources used in this work. We express our gratitude for the assistance provided by Maisy Dunlavy and Matthew Fricke from UNM CARC.

\bibliographystyle{aasjournal}
\bibliography{references}

\appendix

\section{Reconstructing Macrolens Parameters} \label{App:A}

\begin{figure*}
\begin{center}
	\includegraphics[clip, trim=0cm 0cm 0cm 0cm, width=0.9\textwidth]{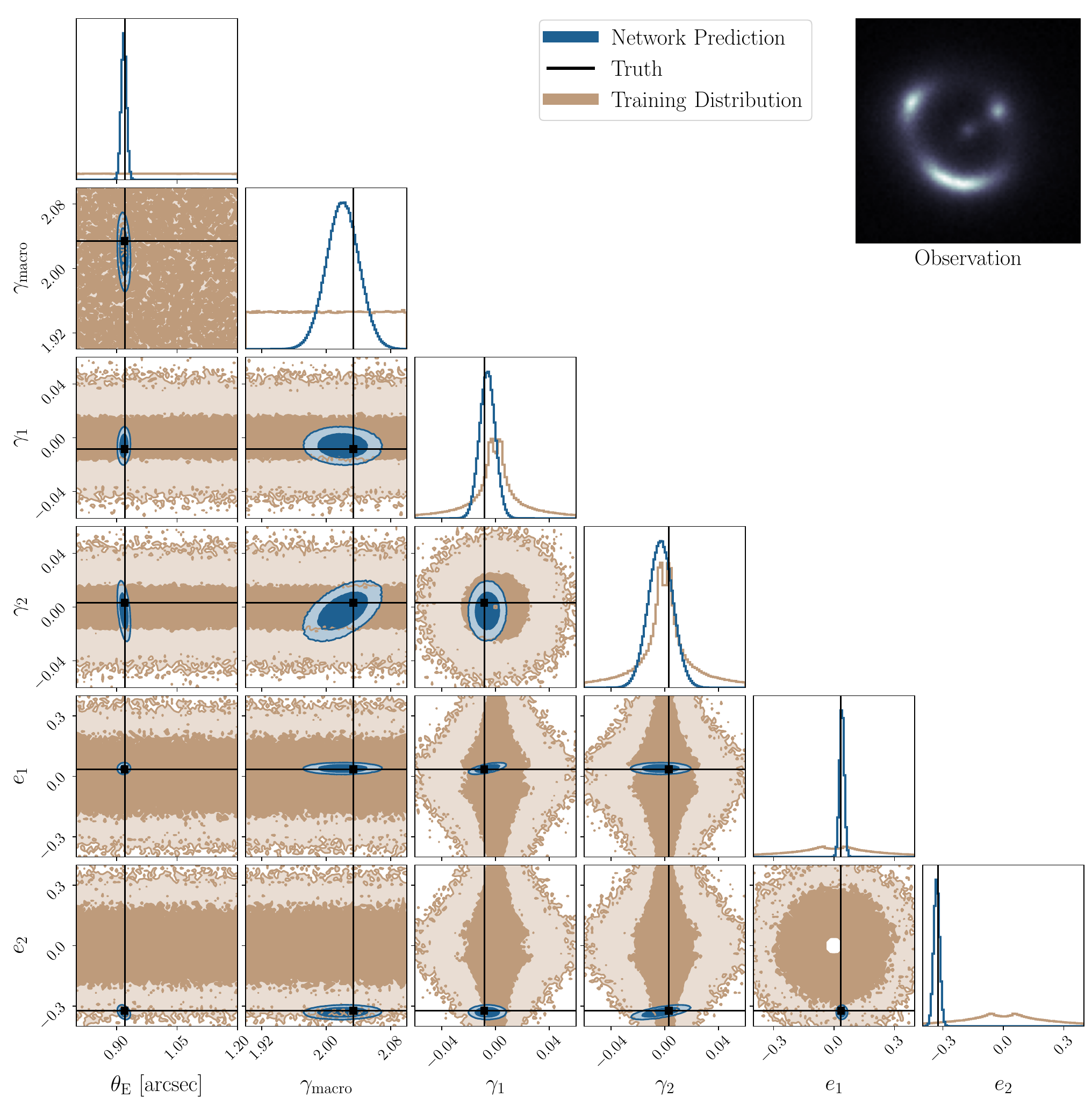}
\end{center}	
\caption{\label{fig:corner_macro} Comparable to Figs.~\ref{fig:dm_params_corner_reandom} and~\ref{fig:corner}, albeit focused on the main lens parameters. This figure showcases the adeptness of the neural density estimator in establishing robust constraints on the main lens parameters.}
\end{figure*}

\begin{figure*}
\begin{center}
	\includegraphics[clip, trim=0cm 0cm 0cm 0cm, width=0.9\textwidth]{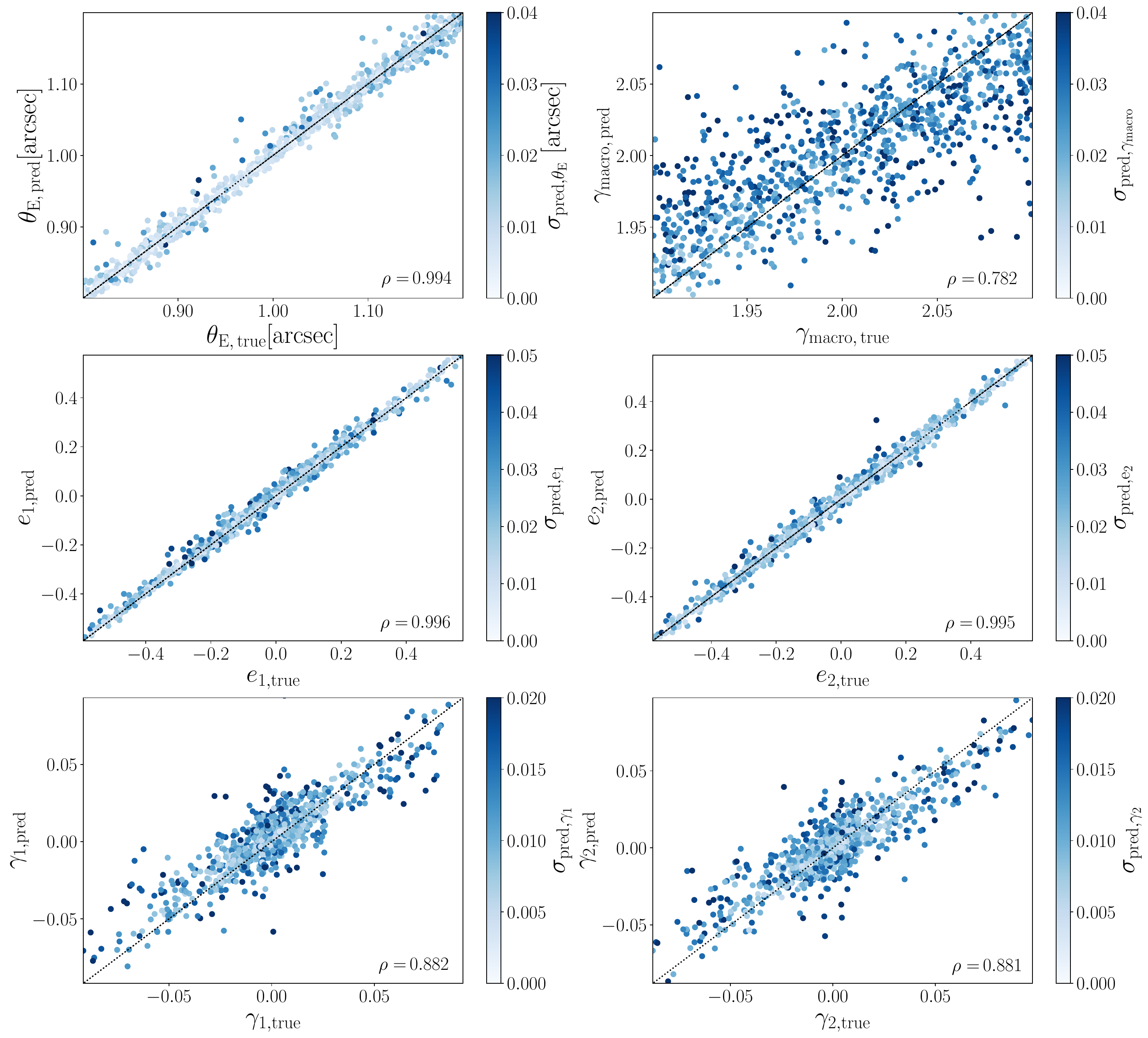}
\end{center}	
\caption{\label{fig:main_lens_true_pred} Similar to Figs.~\ref{fig:dm_params_true_pred} and ~\ref{fig:corr_stat_true_pred}, this figure displays the comparison between predicted and true values, focusing on main lens parameters.}
\end{figure*}

\end{document}